# Masking the general population might attenuate COVID-19 outbreaks

Björn Johansson, M.D., Ph.D.
Theme Aging, Karolinska University Hospital,
Department of Clinical Neuroscience, Karolinska Institutet

Correspondence to:
Email: bjorn.johansson@ki.se
Phone: +46 8 585 80000

**ABSTRACT**
The effect of masking the general population on a COVID-19 epidemic is estimated by computer simulation using two separate state-of-the-art web-based softwares, one of them calibrated for the SARS-CoV-2 virus.
The questions addressed are these:
1. Can mask use by the general population limit the spread of SARS-CoV-2 in a country?
2. What types of masks exist, and how elaborate must a mask be to be effective against COVID-19?
3. Does the mask have to be applied early in an epidemic?
4. A brief general discussion of masks and some possible future research questions regarding masks and SARS-CoV-2.
Results are as follows:
(1) The results indicate that any type of mask, even simple home-made ones, may be effective. Masks use seems to have an effect in lowering new patients even the protective effect of each mask (here dubbed "one-mask protection") is low. Strict adherence to mask use does not appear to be critical. However, increasing the one-mask protection to > 50% was found to be advantageous. Masks seemed able to reduce overflow of capacity, e.g. of intensive care. As the default parameters of the software included another intervention, it seems possible to combine mask and other interventions.
(2) Masks do seem to reduce the number of new cases even if introduced at a late stage in an epidemic. However, early implementation helps reduce the cumulative and total number of cases.
(3) The simulations suggest that it might be possible to eliminate a COVID-19 outbreak by widespread mask use during a limited period.
The results from these simulations are encouraging, but do not necessarily represent the real-life situation, so it is suggested that clinical trials of masks are now carried out while continuously monitoring effects and side-effects.

## INTRODUCTION

Until recently, the World Health Organization recommended that facemasks should be used only by health workers and people with confirmed or suspected coronavirus infection and their carers. [30]. In early 2020, certain news items discouraged the use of face masks (e.g. [15]). However, in China, which was reportedly successful in containing the COVID-19 epidemic, there was a widespread use of facemasks, including by asymptomatic people [13]. Hand hygiene is deemed the cornerstone of infection prevention (see e.g. [1]). However, Jansen discussed the evidence on disinfectants efficacy against SARS-CoV-2 (the virus causing COVID-19) [1]. It was noted that definitive claims on the effectivity of various disinfectants against SARS-CoV-2 could not be made, simply because this is a new virus and the range of off-the-shelf ABHRs has never been tested for SARS-CoV-2. Like other respiratory viruses, the new coronavirus likely spreads from person-to-person through airborne droplets, but it has been pointed out that SARS-CoV-2 might also be transmitted through surfaces, where it can survive for days [40] so that touching infected surfaces can spread the virus (e.g. [1]).

Simple experimental studies suggest that masks may be effective against respiratory infections. For example, Johnson and colleagues had participants cough five times onto a Petri dish containing viral transport medium. Influenza virus could be detected by RT–PCR from all nine volunteers without a mask, no influenza virus could be detected when participants wore surgical or N95 masks (cited in [9]). The same review concluded that there is some evidence to support the wearing of masks or respirators during illness to protect others. Tissue from a surgical mask was found to reduces the risk of COVID-19 transmission in hamsters [6]. Hui et al. [18] found that masks can reduce the distance travelled by expelled air during a cough. Tracht and colleagues noticed that people are willing to wear facemasks to protect themselves against infection [38]. Using mathematical modeling, they concluded that if N95 respirators are 20% effective in reducing susceptibility and infectivity and 10% of the population wear them, the number of H1N1 cases is reduced by 20%.

A variety of masks and related devices exist. Medical masks, unfitted and disposable, can be used by infected individuals, healthcare professionals or laymen to lower transfer of infectious agents [36]. Surgical masks are intended to limit contamination of wounds in surgery. A respirator, a type of mask, is fitted, can be disposable or reusable and protects the wearer against inhalation of harmful material. The National Institute for Occupational Safety and Health (NIOSH) regulates testing and certification of masks and similar respiratory protection equipment [17]. In the European Union, similar standards are provided by the European Committee for Standardization. The NIOSH tests requires a minimum filtration efficiency of 95%, 99% or 99.97% for a test aerosol (see standards for detailed specifications). The more protective masks may offer noticeable resistance to breathing and related to this, some individuals may find them difficult to wear for extended periods. The N95 respirator is a common mask that nominally filters at least 95% of airborne particles [8].

It is indeed possible to obtain very close to 100% protection from respiratory pathogens such as SARS-CoV-2-19. This is using a Self-Contained Breathing Apparatus (SCBA). These are costly and require special training (often used by firefighters), see [16]. An advanced mask (not SCBA) was tested with standardized methods, with influenza, rhinovirus, bacteriophage, *Staphylococcus aureus*, and model pollutants [47]. >99.7% efficiency was found for exclusion of influenza A virus, rhinovirus 14, and *S. aureus* and >99.3% efficiency for paraffin oil and sodium chloride.

An important consideration that is not always mentioned when masks are discussed is the fact that they enable for two barriers to be raised between an infected and an uninfected person. Masks worn simultaneously by infected and uninfected individuals would be expected to compound the reduction of transmission as follows: If the protection of one mask ("one-mask protection") is x, and it is assumed that the size of the protective effect is the same for infected and uninfected persons, the total protection is $(1-(1-x)^2)$, illustrated in Figure 1 (solid line). This would amplify the protection afforded by the masks, and if an infected individual does not wear the mask properly, the masks of uninfected individuals nearby will offer some degree of protection, and vice versa. Notice that masks can be worn by infected individuals to protect other people or by uninfected individuals to avoid respiratory pathogens in the surroundings. Publications do not always make clear the distinction between output protection (from the former situation) and input protection (from the latter situation). Output and input protection may differ, and there is evidence that mask on source is often more effective than mask on receiver [31].

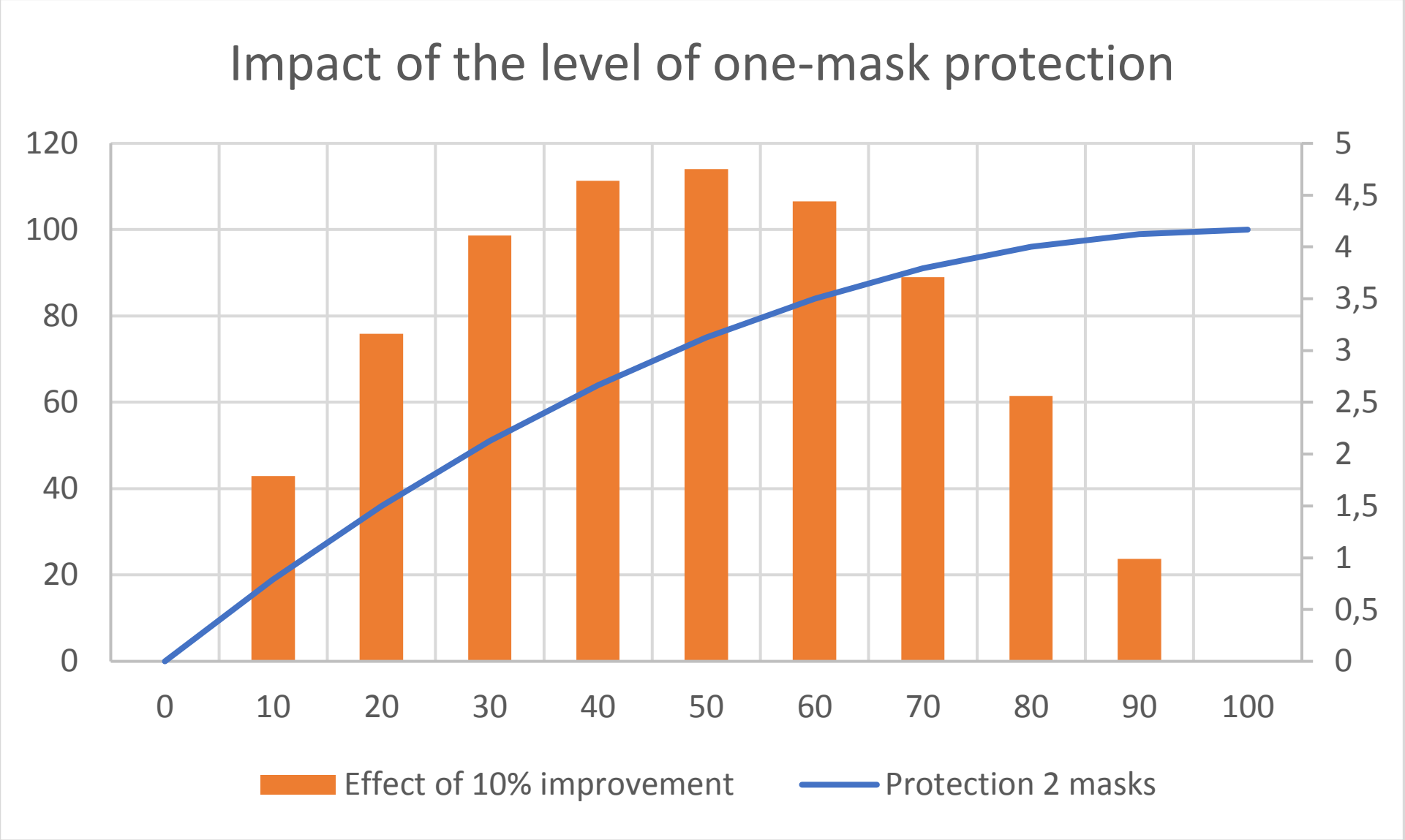

**Figure 1.** Total protection in % (by masking the infected and the healthy individual (blue line, scale to the left) and marginal utility of increased one-mask protection. as a function of one-mask protection. The orange bars with scale to the right show the effect on total protection of a 10% increase in one-mask protection.

## Transmission of respiratory pathogens through masks. Different ways to measure transmission and protection

The protective ability of a mask can be expressed and measured in different ways, not always discussed in a comprehensive manner. The share of virions (or other particles) that pass through a mask is often termed "penetration" and determined as the ratio between the concentrations inside and outside the mask. The "efficiency" of a mask is a measure of how much of the agent is turned away be the mask and is 100% - penetration [42].

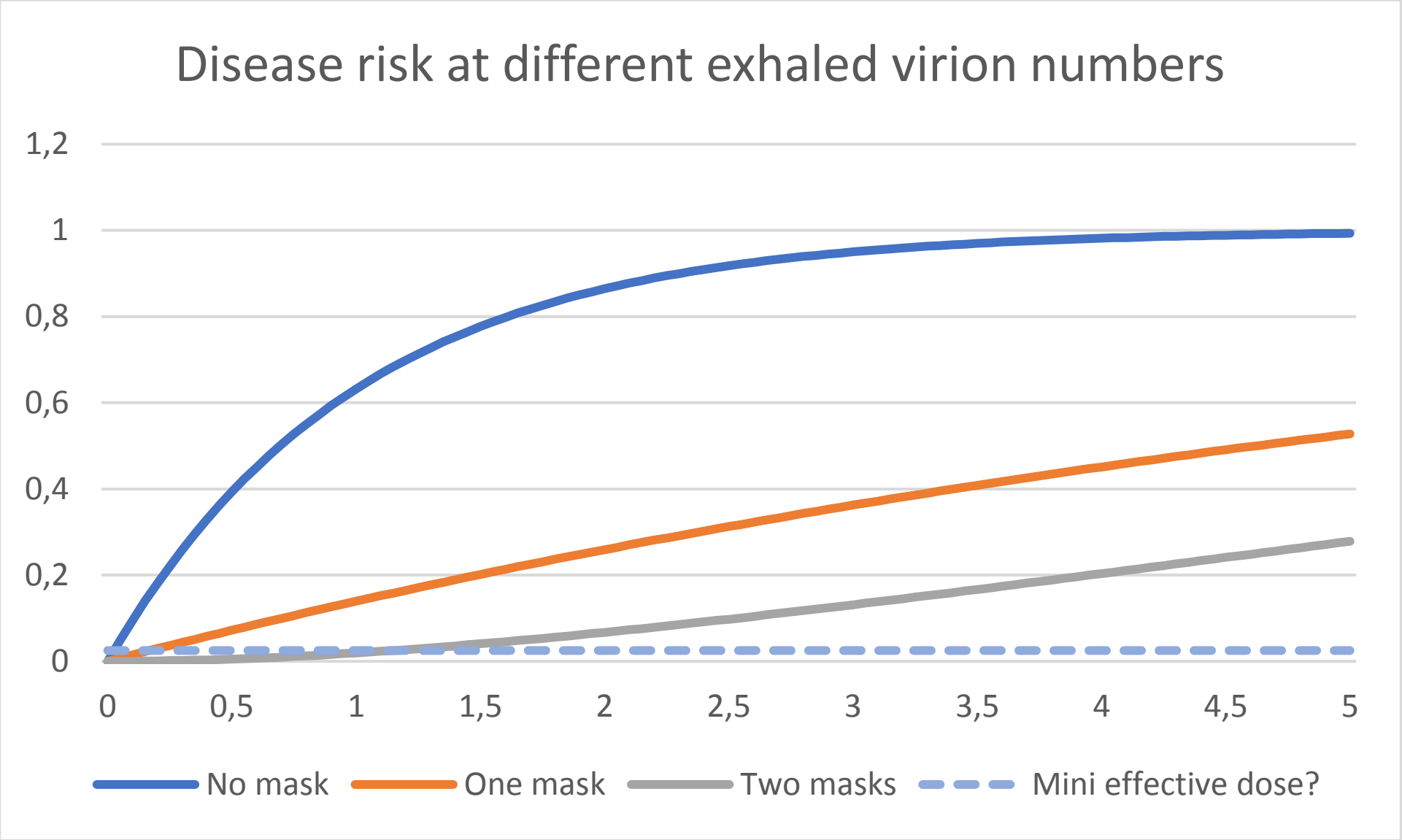

**Figure 2.** A sketch that attempts to illustrate the transmission of respiratory pathogens through masks worn by infected virus sources and uninfected recipients. The units on the horizontal axis are arbitrary. The vertical axis shows the likelihood of being infected.

Notice that at very high concentrations of virions (to the right in Figure 2), there is so much excess of virions received that the number of virions received becomes less important and therefore the risk reduction by one or even two masks becomes small. On the other hand, there might be a minimal infective dose of virions, below which infection does not occur. This could mean that a small reduction in virion count by mask use might completely abolish infection. However, the existence of a minimal infective dose is under debate, may be situation dependent, and for some viruses, the minimal infective dose may be equal to 1 [45]. Of possible relevance here is the curious observation that a high percentage of morphologically identical viral particles in a sample, as determined by electron microscopy, will often be non-infectious. A dose-response (illness) curve has been published for SARS [41].

### Spread of respiratory infections: droplets, aerosols and other routes

Several routes and modes exist for transmission of respiratory infectious diseases; droplets may contribute to several of them. The modeling by Stilianakis and colleagues divided into respirable droplets, with droplet diameter less than 10 µm, and inspirable droplets, with diameter in the range 10–100 µm. According to these authors, droplet dynamics is determined by their size, whereas population dynamics is determined by, i.a., pathogen infectivity and host contact rates. Robinson et al. of the team just mentioned [35] suggested that small droplets (~0.4µm) have too small viral load to be significantly infection and that larger droplets (~4µm) are the primary vehicle for infection. For SARS-CoV-2, a recent publication [46] argues from global trends in the number of infected individuals that airborne transmission is the dominant route, among several. Tellier et al [37] has argued that SARS-CoV and MERS-CoV (viruses that cause SARS and MERS) may have to penetrate directly into the lower respiratory tract before causing disease; these authors also notic that terminology in this area is not uniform. Whether an infectious agent is transferred by large droplets or airborne/aerosol may be important to the choice of protective equipment, as there is evidence that a conventional surgical mask insufficient to protect against aerosol transmission, and that more elaborate masks may be more appropriate.

**Estimates of the protection offered by masks**

It was found that under unfavorable conditions, more than 3% of MS2 virions penetrated through filters of N99 and N95 respirators [2]. Wiwanitkit and collaborators [43] found that the size of the pores of the N95 mask is about 300-500 nm in diameter whereas the size of the avian flu virus is about 100 nm (SARS size may be similar). I.e. 3–5 times larger than virus and there is evidence that SARS can pass through N95s [4]. Simple fabrics were reported to have 40-90% instantaneous penetration levels of polydisperse NaCl aerosols, much worse than for N95 respirator filter media. In addition, N95 masks also has about a 10% leakage problem around the mask. The study by Rengasamy et al. [34], suggests that the upper level of efficiency for the common N95 mask may be 85% - i.e. 10% leakage on the side and 5% penetration through the filter.

As the great majority of studies on masks have been carried out with other pathogens than SARS-CoV-2, a pertinent question is whether the size of the protection by mask is similar for different respiratory pathogens or not. This is addressed in the work of Zhou et al. [47] indicates that the efficiencies of masks for excluding different pathogens may be similar. Eninger [11] concluded studying different masks that measuring the penetration of simple NaCl aerosols may generally be appropriate for modeling filter penetration by virions.

Balazy et al. found evidence that different models of the same type of mask can have very different protective properties [2]. Zuo et al. [48] found that although physical penetration of adenovirus and influenza virus aerosols through respirators can be substantial, 2%-5%, infectivity penetration of adenovirus was much lower. In the meta-analysis of Offeddu et al. [29] quantified the protective effect of facemasks and respirators against clinical respiratory illness (risk ratio [RR] = 0.59) and influenza-like illness (RR = 0.34). Meta-analysis of observational studies provided evidence of a protective effect of masks (odds ratio OR = 0.13) and respirators (OR = 0.12) against severe acute respiratory syndrome (SARS).

## Simulations of early and late interventions

This study attempts to estimate the effect of masking the general population on a COVID-19 outbreak first using simple considerations about the basic reproduction number of the epidemic and the level of protection from different types of masks. The basic reproduction number ($R_0$), is an index of the contagiousness of an infection and depends on both the infectious agent and other factors. As $R_0$ is the expected number of secondary infections produced by an index case in a completely susceptible population, it is often used to predict if an outbreak is expected to continue, as $R_0 > 1$ indicates that it will and $R_0$ is $< 1$ indicates that it won´t. However, this is a simplification and the calculations surrounding $R_0$ can be complex [10] [44]. This study begins by gathering some published numbers regarding $R_0$ of the COVID-19 epidemic and the protection afforded by masks. This study then moves om to simulations with two separate web-based softwares, one of them a specialized COVID-19 program and one a simpler program intended for educational purposes.

Then the effect of masking is addressed using computer simulations. Claims are sometimes made that disease preventive measures have to be enacted early in order to be effective. Early intervention has been claimed to be important also regarding COVID-19 (for example [19] [32]). We therefore tested early and late intervention using the computer modeling.

## MATERIALS AND METHODS

This study used software developed by others for computer simulations of a standard population level epidemiological models. As described in detail [28], COVID-19 Scenarios simulates a COVID-19 outbreak with a generalized SEIR model with the total population divided into age-strata (because of known age dependence of COVID-19 outcome) compartments of: susceptible (S), exposed (E), infected (I), hospitalized (H), critical (C), ICU overflow (O), dead (D) and recovered (R) individuals. People transition among the different compartments. The model allows to specify individual interventions with start and end dates to model the existing (social distancing, case isolation and quarantine) as well as additional interventions. COVID-19 Scenarios provides default parameters estimated from real-life statistics, although the authors emphasize the uncertainty behind these estimates. We ran the simulation for the United States. The model had been calibrated by its authors to match its age structure and the observed epidemiological statistics. The simulations did not consider saturation phenomena that might occur at very high virus counts (Figure 2), neither a possible minimal infectious dose of virions (Figure 2). It was assumed that transmission occurred only through routes that can be blocked by mask use.

**Table 1.** Parameters used for COVID-19 Scenarios. These were the default parameters provided by the software when the simulations were done. Please notice that the defaults include a 73.8%-84.2% intervention introduced on March 24, that is included with the software. When mask interventions were included in the simulation, they were introduced on January 1 (i.e. for the whole period of the simulation) or July 1.

## Scenario: United States of America (edited)

## Parameters

### Population

| Parameter | Value |
|---|---|
| ageDistributionName | United States of America |
| caseCountsName | United States of America |
| Number of hospital beds | 798288 |
| icuBeds | 49499 |
| Cases imported into community per day | 0.1 |
| Number of cases at the start of the simulation | 1 |
| Population size | 327167434 |

### Epidemiology

| Parameter | Value |
|---|---|
| hospitalStayDays | 3 |
| icuStayDays | 14 |
| infectiousPeriodDays | 3 |
| latencyDays | 3 |
| Increase in death rate when ICUs are overcrowded | 2 |
| Seasonal peak in transmissibility | January |
| R0 at the beginning of the outbreak | 4.1 - 5 |
| Seasonal variation in transmissibility | 0 |

### Mitigation

| Intervention name | From | To | Reduction of transmission |
|---|---|---|---|
| Intervention 1 | Mar 24 2020 | Sep 01 2020 | 73.8% - 84.2% |

**Table 2.** Four cases, representing four different levels of protection by mask, that have been considered in this study and incorporated into the simulations that use COVID-19 Scenarios.

| Case no. | Pathogen removal/risk reduction, one mask | Removal/reduction, two masks | References |
|---|---|---|---|
| 1 | 99.7% | 99.9991% | [47]. An advanced mask. |
| 2 | 85% | 97.75% | N95 mask. This is estimated from data in [24], [31] assuming 5% filter penetrance and 10% leakage on the sides. |
| 3 | 22% | 39,16% | Average of input and output protection by a simple home-made mask; based on measurements in van der Sande et al. [8]. |
| 4 | 5.7% | 11% | Based on the 0.89 relative risk reported in [1], which is a meta-analysis of Hajj pilgrims. Notice that in some segments of the population studied, actual mask use was <<50%. |
| | | | The question to be addressed here is whether masks can influence the epidemic even if many don´t use their masks properly. |

Notice that although most of the numbers in Table 2 are taken from published papers, they need not be representative for all pathogens or varieties of masks. None of the numbers in the table derives from a study with COVID-19.

The second software used in this paper is Epidemix [26], which is a simplified software for teaching and demonstration purposes. It uses a visual interface to access eight models of epidemics without dealing with the details of mathematical equations and program code. The underlying calculations are done by a set of software packages in the programming language R. All models simulate disease spread through a population, allowing the user to select the model and characteristics of the population, interventions etc.

Curve-fitting to estimate the slope of the curve of scenarios with different late mask interventions was done using the diagram function in Excel for Microsoft 365.

## RESULTS

### 1. A simple estimation of the degree of protection afforded by masks and the protection needed to influence the epidemic

The transmissibility of a virus is measured by the basic reproduction number ($R_0$), which measures the average number of new cases generated per typical infectious case. As described by Rahman et al. ([33] and references therein, $R_0$ of 1.0 is an important threshold value. If $R_0$ is equal to 1 or less, this indicates that the number of secondary cases will decrease over time and, eventually, the outbreak will peter out. One review evaluated the mean and median of $R_0$ estimated by the 12 articles and they calculated a final mean and median value of $R_0$ for COVID-19 of 3.28 and 2.79 [22]. This seemingly provides a rough indication of how much the transmission must be reduced to reverse the epidemic. It appears that a reduction of transmission of at least two thirds is necessary. This is within the range of protection of some but not all masks (some 67% efficiency). However, as there would be two barriers between infected and non-infected individuals, the numbers for two serially connected masks should presumably be calculated with, as the mask of the uninfected individual will add to the protection from the mask of the infected individual, indicating that many of the available masks might be adequate.

## 2. Simulations using COVID-19 Scenarios modeling masks during the start of an outbreak

Figure 3 shows the simulated number of confirmed COVID-19 cases in the U.S. vs. date. The vertical axis is a logarithmic scale. The linearity of this graph shows the exponential growth expected early in an epidemic. We ran the simulation for the period ending with August 31, 2020. This was long enough to see effects of masks on the time course of the epidemic. We wanted to observe the sensitivity of the outcomes to the time when mask intervention was begun, rather than the effects of different mitigations implemented at the same time.

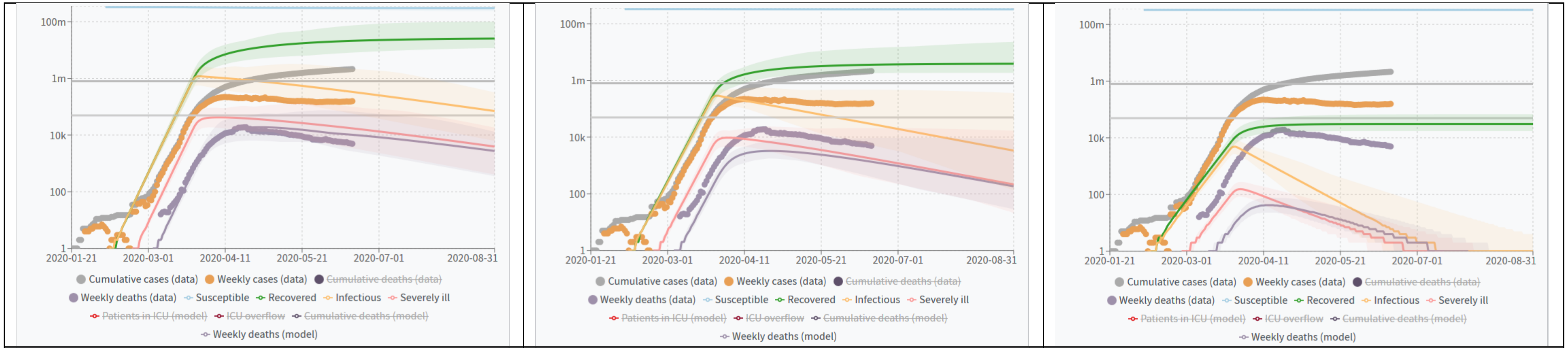

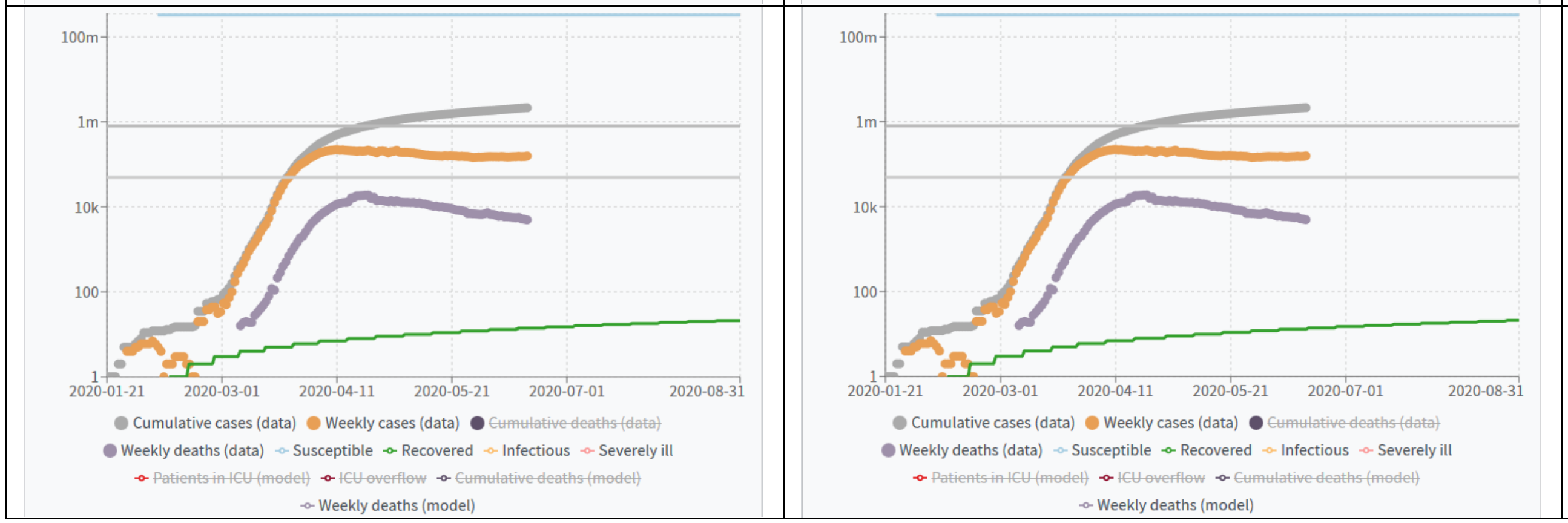

**Figure 3.** The effect of masking the general public on the COVID-19 epidemic, as estimated using simulation by Covid-19 Scenarios. Upper row, no mask, 11% protection and 39.% protection. Lower row, 97.75 and 99.9991% protection.

| time 2020-08-31 | cumulative recovered (total) median | cumulative recovered (total) lower bound | cumulative recovered (total) upper bound | cumulative hospitalized (total) median | cumulative hospitalized (total) lower bound | cumulative hospitalized (total) upper bound | cumulative ICU (total) median | cumulative ICU (total) lower bound | cumulative ICU (total) upper bound | cumulative fatality (total) median | cumulative fatality (total) lower bound | cumulative fatality (total) upper bound |
|---|---|---|---|---|---|---|---|---|---|---|---|---|
| Default | 25435572 | 11699016 | 100463893 | 892376 | 410277 | 3528347 | 468735 | 218699 | 1754633 | 224006 | 97251 | 996242 |
| Default + mask 11% (scenario 4) | 3879172 | 1071446 | 10906385 | 136042 | 37572 | 382708 | 73074 | 20270 | 200586 | 33934 | 9498 | 89024 |
| Default + mask 39% (scenario 3) | 30719 | 17446 | 67805 | 1078 | 612 | 2378 | 581 | 330 | 1282 | 273 | 155 | 600 |
| Default + mask 97.5% (scenario 2) | 21 | 21 | 21 | 1 | 1 | 1 | 1 | 1 | 1 | 0 | 0 | 0 |
| Default + mask 99.9991% (scenario 1) | 21 | 21 | 21 | 1 | 1 | 1 | 1 | 1 | 1 | 0 | 0 | 0 |

**Table 3.** Data from COVID-19 Scenarios regarding number of individuals in the different compartments of the model.

The results from the first runs with COVID-19 Scenarios show that mask use appears to be effective even at low one-mask protection or limited compliance. Even the lowest efficiency scenario reduced the simulated epidemic if applied from its beginning. The two highest protective abilities seem to practically completely abolish the epidemic. There was no indication of any difference between the two masks with the highest protection modeled (scenarios 1 and 2). It appears that even simple masks (e.g. 21% protection, scenario 2) or low-compliance mask-wearing (11% protection, scenario 1) reduces the number of COVID-19 cases within weeks.

### 3. Simulations using Epidemix version 2

These simulations used the full defaults of Epidemix version 2, i.e. the parameters were not specifically calibrated for SARS-CoV-2 (deterministic homogenous model, infection states S, Is, R, population size 100, daily number of effective contacts per unit 0.4, length of infectious period 10 days). An added intervention with 75% two-mask protection, corresponding to a simple mask, was tested by reducing transmission by 75%.

**Figure 4.** Simulations using Epidemix version 2 (upper): 10% infected initially, otherwise full defaults of the Epidemix program. (lower) 10% infected initially, 75% two-mask protection, otherwise full defaults of the Epidemix software.

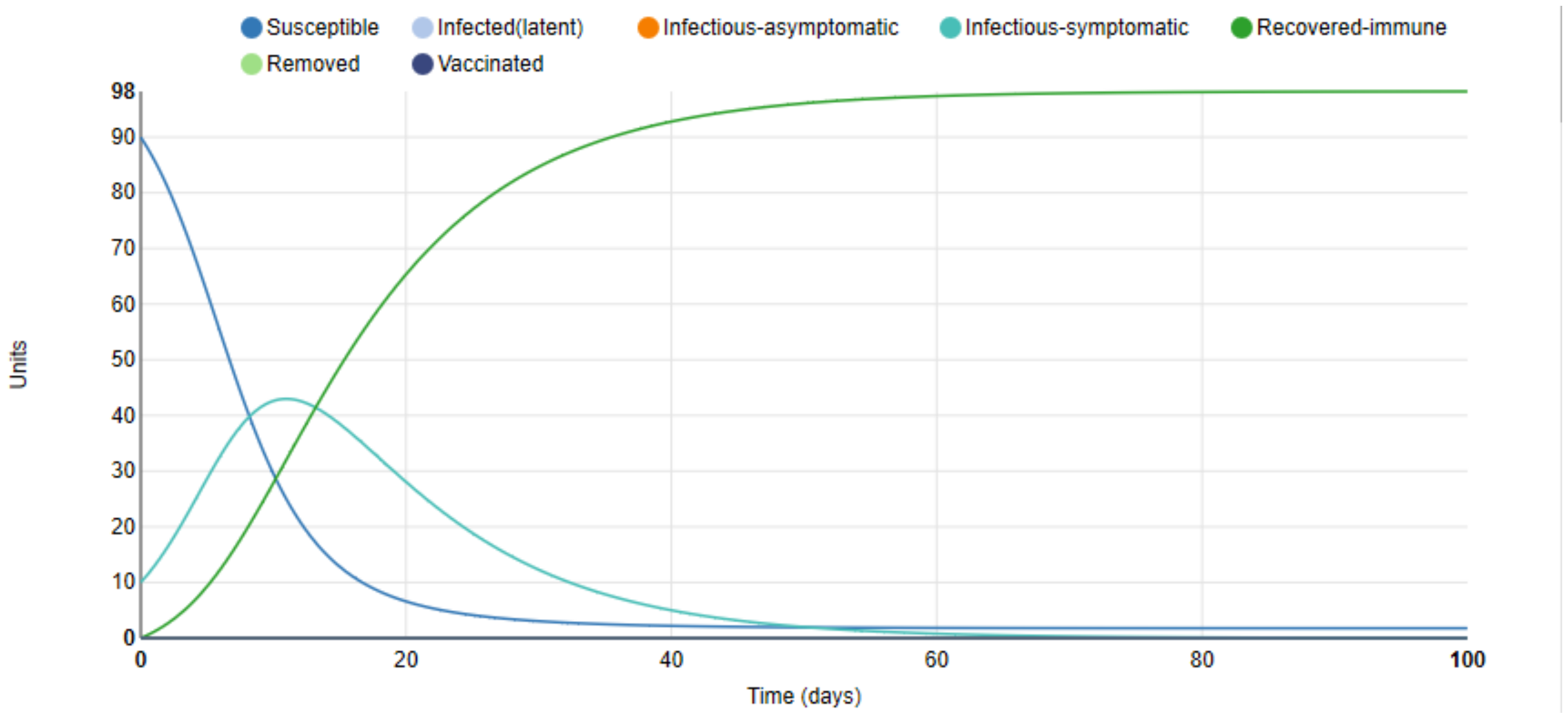

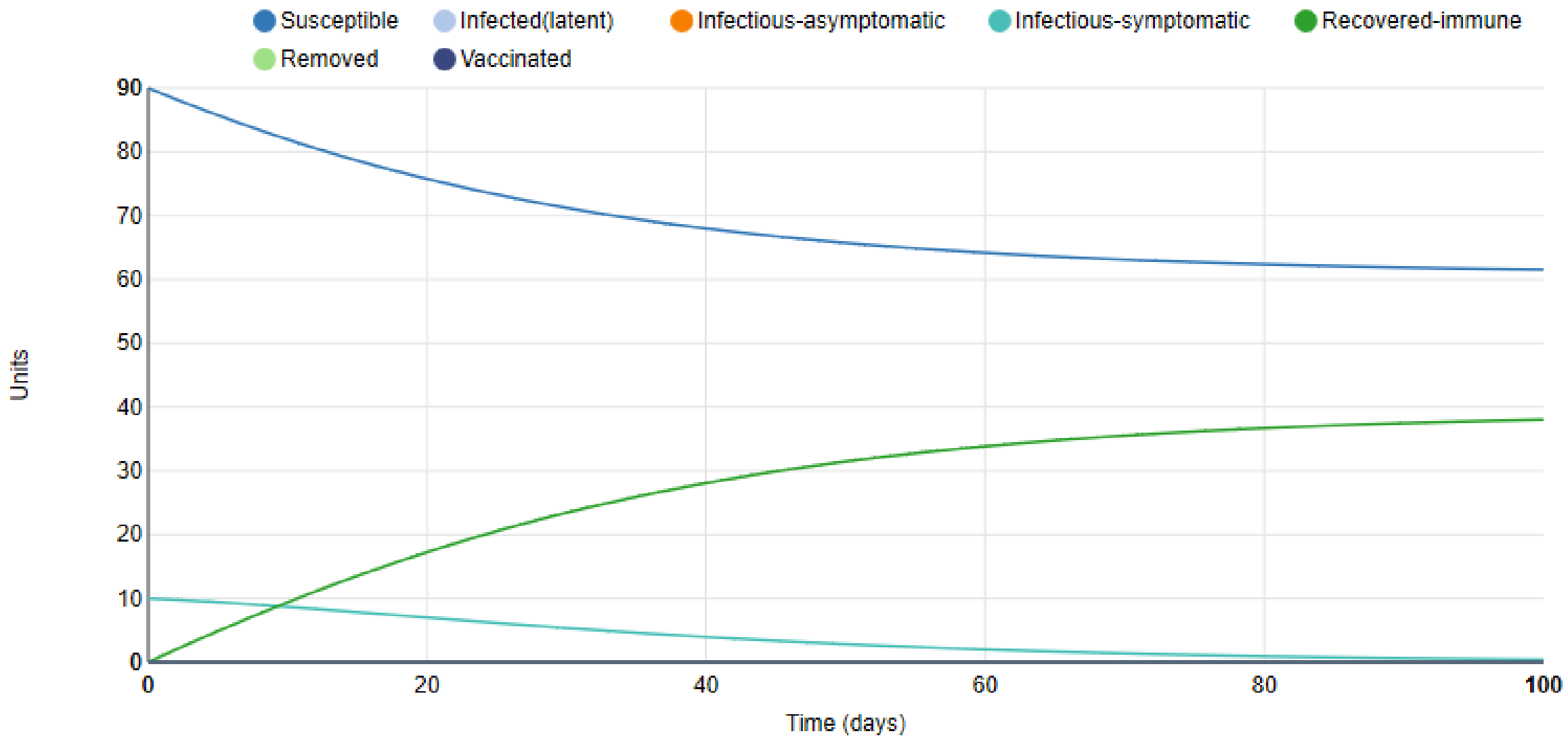

As shown in Figure 4, the cumulative number of infected individuals during the whole simulation was only half in the mask use scenario as compared to the default scenario. with the effect of mask present throughout the period. The result seemed to corroborate the result from the COVID-19 scenarios model that masks reduce COVID-19 cases.

### 3. Do interventions have to be applied early?

From the results in Figure 5, it seems that they don´t. When they are applied late, the total number of cases is influenced less than active cases, since a mask will not help individuals who have already been infected. Hover, the number of active cases is reduced with all mask interventions modeled, with some interventions resulting in a dramatic reduction. With case 4, 11% two-mask protection, reduction in new cases was about 11% per week. With case 1, reduction was quite dramatically approx. 88% per week.

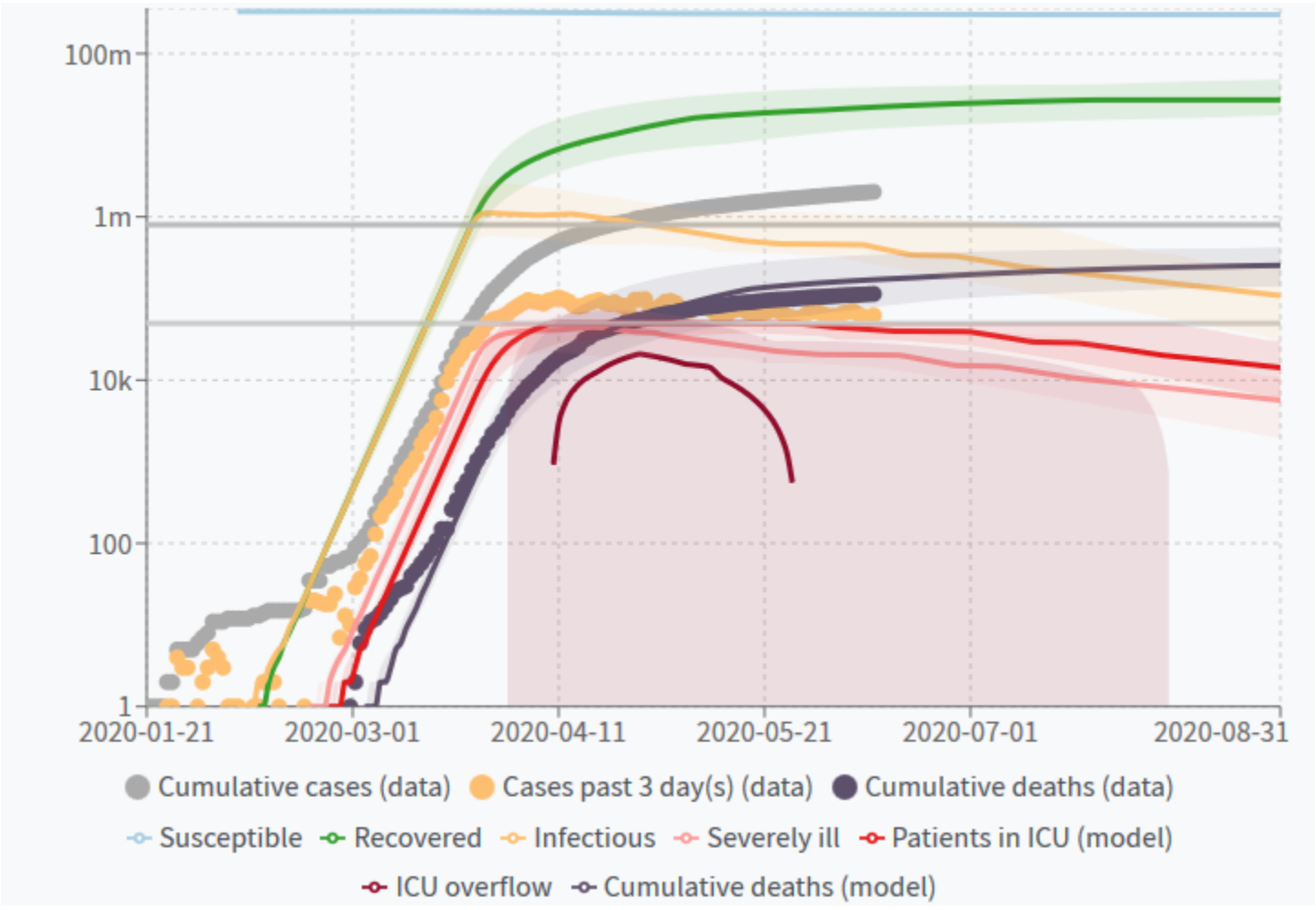

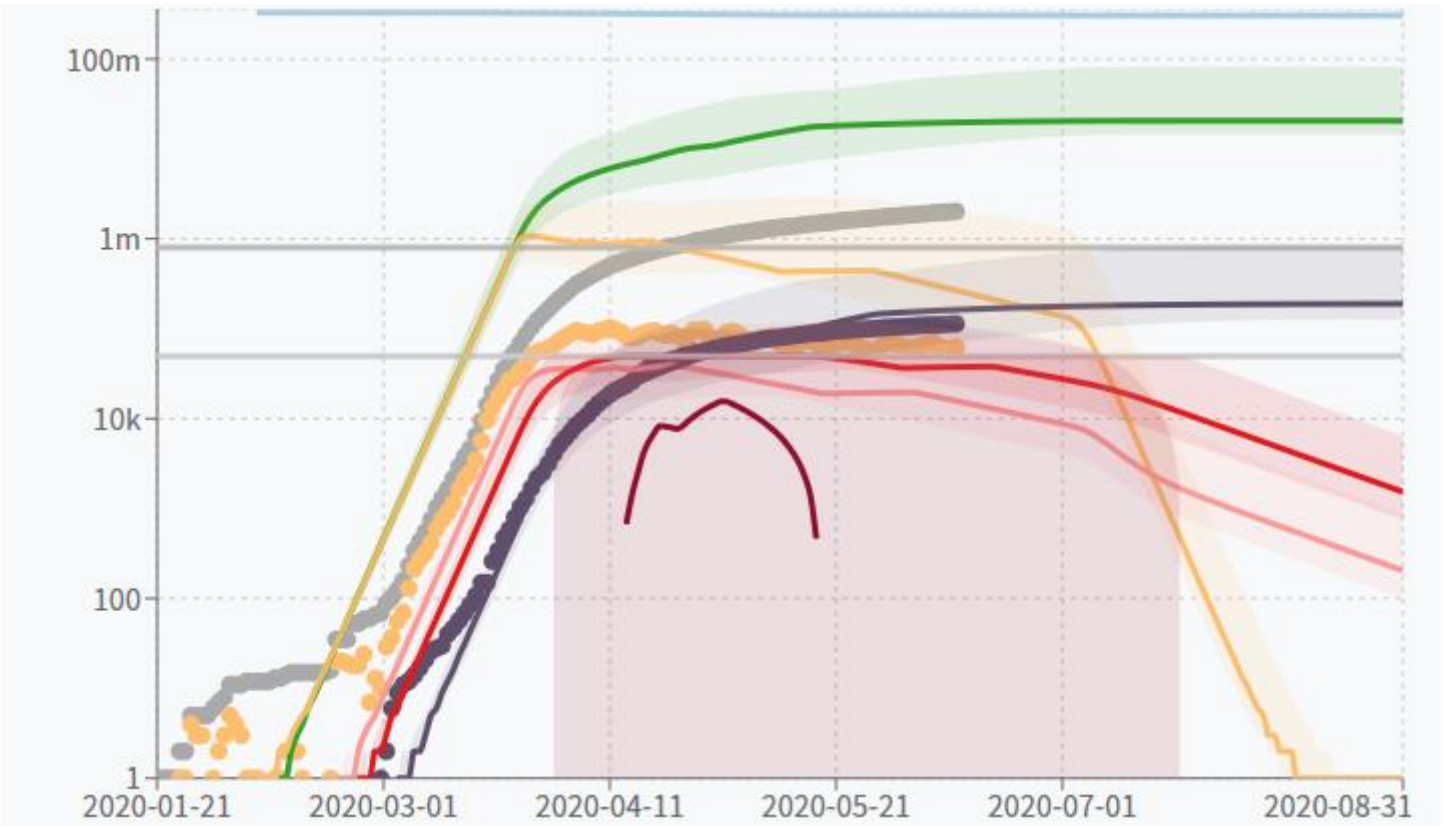

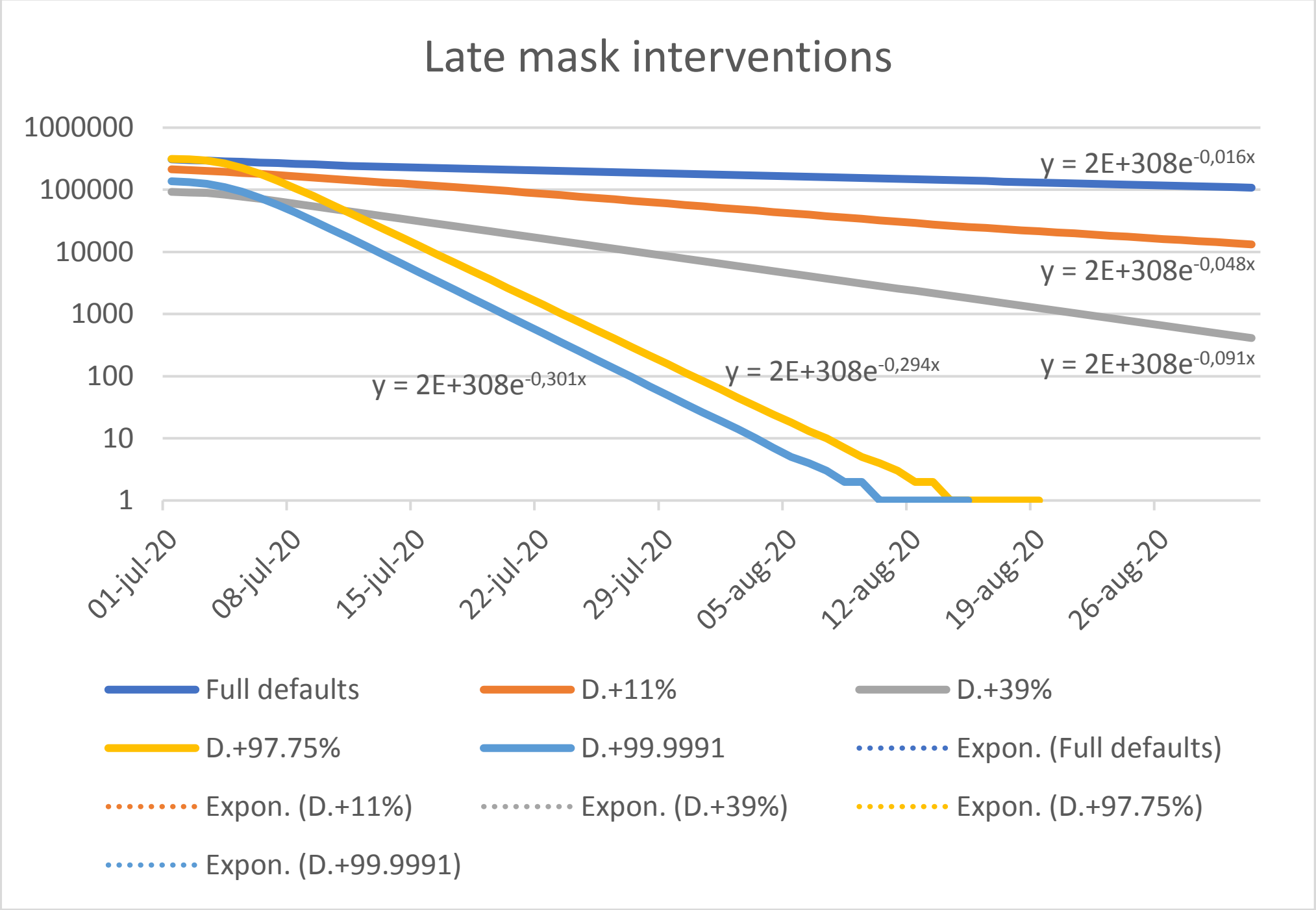

**Figure 5.** Effects of late mask interventions on the number of active COVID-19 cases. (Upper graph) Full defaults according to Covid-19 Scenarios. (Middle graph) Full defaults with late mask intervention using advanced mask (99.9991% two-mask protection). (Lower graph) Estimating half-life and weekly reduction from the late use of four different masks (case 1-4); vertical axis is "Infectious" individuals from COVID-19 Scenarios.

**Table 4.** Estimates of $T_{1/2}$ of "Infectious" individuals with different late mask interventions.

| Multiplier in exponent estimated from graphs | Half-life | Weekly reduction in active COVID-19 cases (%) |
|---|---|---|
| -0,016 | 43,32169 | 11 |
| -0,048 | 14,44056 | 29 |
| -0,091 | 7,617 | 47 |
| -0,294 | 2,357643 | 87 |
| -0,301 | 2,302814 | 88 |

## DISCUSSION

### Even small protective effects from masks may compound over time

It was found that even the smallest mask intervention had an effect long-term in the simulations. Even with just 11% protection (the low compliance case), there was an effect on the numbers of active infections. If intensive care units are working close to capacity, this could mean that even adoption of the simplest mask could result in a large reduction in deaths. Furthermore, this effect was found to compound over time so that after several weeks there is a large decrease in the number of new COVID-19 cases. This is reminiscent of interest-on-interest in a bank account, when the interest becomes sizable after applied repeatedly. With mask protection > 50% protection was simulated, the effect on the size of the epidemic was dramatic, from about 1 million to zero fatalities (Table 3). It therefore appears that wearing of N95 masks (85% protection assumed) by the population can dramatically reduce the number of COVID-19 cases. There was no different between the two mask interventions with the highest protection factor.

### A "corona washout" may be possible in a limited number of weeks

The calculations illustrated in Figure 5 and Table 4 suggest that complete elimination of COVID-19 can be achieved in a closed community with less than 2 months of intervention with highly protective masks. It would be of interest to identify a community suitable for a clinical trial with such masks. Any such study should be carefully carried out taking local conditions (incl. legislation) into account and using continuous evaluation of infection parameters as well as any side-effects of the masks.

It should be said that the results above seem to be robust. I.e. the principal results do not seem to depend on precise model used or its input parameters, as slight variations of some parameters that have been tested have not altered the fundamental results (unpublished results).

**Side effects and risks of masks**

There may be few comprehensive studies of side effects of masks. From everyday experience, the side-effects and risks of mask use are usually limited to minor discomfort. However, serious side-effects from some masks may be possible: One of the more serious side-effects of mask-wearing seems to be that some masks may cause increased condensation on eyeglasses worn concurrently [25], potentially obstructing the view of the bearer. Bacteria can survive on surfaces of masks for several days [4]. Personal protective equipment was reported to be a source of airborne infections [27]. Pain and pressure from masks have been described during COVID-19 pandemic, and remedies suggested [12]. Gefen et al. studied device-related pressure ulcers in the context of COVID-19 [14]. Ju et al. reported contact vitiligo from rubber ear loops from a mask [20]. Side effects of respirators include symptoms related to hypercapnia [49], work of breathing and gas exchange [3]. Finally it could be mentioned that heavy respiratory protective equipment used by firefighters can affect the balance of the user [5], potentially increasing the risk of fall accidents, although this may be less relevant to most masks discussed in this paper. This paragraph does not attempt to be a complete list of possible side-effects of masks.

**Questions for additional research**

The conclusion is that masks should be evaluated as an important addition to other ways of protection. They may have protective effects on the same order of magnitude as vaccines, but with the added advantages of being effective against a wide range of respiratory pathogens and can be prepared in advance and stored.

*Questions that can be addressed in future research on masks* is first and foremost a systematic study of side effects and risks associated with masks. Other research on masks would be to be a cost-benefit analysis to decide what at what level of protection is best when costs and side effects are considered. The orange bars in Figure 1 show the marginal improvement in two-mask protection from a 10% improvement in one-mask protection and indicates that the marginal benefit may be biggest from raising protection above 50% and less benefit may be received from improvements in one-mask protection beyond 90%. It may be valuable to study masks with aerosols for specific illnesses before an airborne epidemic hits next time. Other research on masks would be to evaluate the shelf-life of different materials used in masks to select materials that allow for long-term storage preparation for future epidemics. Procedures for industrial and home manufacturing could be optimized. How to best educate individuals about the value of masks, how to properly wear a mask and perhaps how to make their own mask and how to increase adherence to mask use?

*Questions that can be addressed regarding interactions with other interventions.* E.g. do masking and social distancing work synergistically or do they work best on their own? In the simulations above, it implied that masks and other interventions depress SARS-CoV-2 transmission independently, but this is not necessarily the case in real life. Does the introduction of masks reduce people´s compliance with other preventive measures, or do masks serve as a reminder of the epidemic, improving compliance with other measures? Can groups with low vaccine response (e.g. possibly those in [23]) be identified and should they wear a mask instead?

*Questions that can be addressed in future research on SARS-CoV-2* (and other pathogens) would be to confirm and quantify different ways of transmission. Such information would inform decisions about whether a mask is of any value and in which situation it may be of value.

The results from these simulations are encouraging, but the only way to be sure about the effects of masks is to conduct prospective, controlled studies; perhaps along the lines of Lin et al. [21]. Also, masks and related equipment are associated with significant side effects and risks that should be carefully monitored during any implementation. The numbers used here are consistent with the literature but do not represent the whole literature and many numbers derive from study of other agents than SARS-CoV-2. What tells us that the work described here may still be relevant for SARS-CoV-2 is that trying slightly different input parameters or slightly different models typically has not changed the outcome of a simulation much. While this work was ongoing, several studies came out in favor of mask use by the public (e.g. [7] [27] [39] [46]).

**This is an early, preliminary version that has not been submitted for journal publication. The intention is to submit a more complete, corrected version of this manuscript for peer-reviewed publication. This early version is made available in the interest of rapid information sharing among professional researchers. This manuscript is not health, medical, legal or other professional advice. Members of the public should seek the advice of a qualified health professional and the relevant authorities before attempting or disapplying any procedure or using or disusing any equipment mentioned or implied in this manuscript. This work rests on many assumptions and simplifications that may or may not turn out to be valid. Any action taken upon this work is strictly at your own risk. Any information provided is "as is", without any guarantee of completeness or accuracy or of any results obtained. Any views expressed here do not necessarily reflect the official policy or position of any agency, organization, employer or company.**